\documentstyle[12pt]{article}
\title{On an integrable reduction of the Dirac equation}
\author{R.Z.~Zhdanov \\ \small Institute of Mathematics,\\
\small 3 Tereshchenkivska Street,
252004 Kiev, Ukraine\thanks{e-mail: rzhdanov@apmat.freenet.kiev.ua}}
\date{}
\newtheorem{theo}{Theorem}
\newtheorem{lemma}{Lemma}
\begin{document}
\maketitle
\begin{abstract}
A symmetry reduction of the Dirac equation is shown
to yield the system of ordinary differential equations
whose integrability by quadratures is closely connected
to the stationary mKdV hierarchy.
\end{abstract}

Consider the Dirac equation of an electron
\begin{equation}
\label{dirac}
i\sum_{\mu=0}^3\,\gamma_\mu\psi_{x_\mu} -
\left(e\sum_{\mu=0}^3\,\gamma_\mu A^\mu + m\right)\psi = 0,
\end{equation}
moving in the electric field
\[
A_0=A_0(x_3),\quad A_1=A_2=A_3=0.
\]
In the above formulae $\gamma_\mu$ are $4\times 4$ Dirac matrices,\
$\psi=\psi(x_0,x_1,x_2,x_3)$ is a four-component complex-valued function
and $e, m$ are constants.

The form of the vector-potential $A_\mu$ imply the following Ansatz
for the spinor field $\psi(x)$:
\[
\psi(x)=\varphi(x_3).
\]
Inserting this expression into the Dirac equation (\ref{dirac}) yields
system of ordinary differential equations (ODEs) for the four-component
function $\varphi(x_3)$
\begin{equation}
\label{ode1}
\varphi^\prime - (ie\gamma_3\gamma_0A_0 - im\gamma_3)\varphi = 0.
\end{equation}

Denoting
\begin{eqnarray*}
&&x=2x_3,\quad V(x)=eA_0(x_3),\\
&&J_1=\frac{1}{2}\gamma_0,\quad J_2=\frac{i}{2}\gamma_3,\quad
J_3=\frac{i}{2}\gamma_3\gamma_0,
\end{eqnarray*}
we rewrite (\ref{ode1}) in the following form:
\begin{equation}
\label{ode2}
{\cal L}\, \varphi\equiv (D_{x} - V(x)J_3 - mJ_2)\varphi=0,\quad
D_x=\frac{d}{dx}.
\end{equation}

Note that the $4\times 4$ matrices $J_1, J_2, J_3$ fulfill the commutation
relations of the Lie algebra $so(3)$
\begin{equation}
\label{com}
[J_a,\ J_b]=J_c,\ \ (a,b,c)={\rm cycle}\ (1,2,3).
\end{equation}

To integrate system of ODEs (\ref{ode2}) we will make use of its
symmetry properties. The general routine for calculating symmetry
group admitted by a differential equation is the infinitesimal Lie
method. This method makes it possible to reduce the problem of
constructing the maximal symmetry group to integrating some linear
system of partial differential equations (called determining
equations). The general solution of the latter gives rise to the
maximal transformation group admitted by the equation under study (for
more detail, see, e.g. \cite{olv,ovs}). However direct application of
the Lie method to system (\ref{ode2}) is inefficient, since
integration of the corresponding determining equations is, in fact,
equivalent to integration of the initial system of ODEs (\ref{ode2}).

That is why to be able to integrate the determining equations one has inevitably
to impose some {\em a priori} restrictions on the choice of symmetry operators.
Our idea is to look for a symmetry operator $Q$ admitted by system (\ref{ode2})
in the form of an $n$th order polynomial in $m$ with matrix coefficients.
First, we consider in some detail the case when $n=3$ and then give the
results on calculating Lie symmetries obtained for the case of an arbitrary
$n\in {\bf N}$.

Thus, we adopt for a symmetry operator the following Ansatz:
\begin{equation}
\label{symm}
Q=\sum_{k=1}^3\,(a_k(x)+ b_k(x)\, m + c_k(x)\, m^2 + d_k(x)\, m^3)J_k,
\end{equation}
where $a_k,\ b_k,\ c_k,\ d_k$ are some smooth complex-valued functions.

Inserting the expression for $Q$ into the invariance criterion
$[{\cal L},\ Q]=0$ and splitting with respect to the powers of $m$
and then with respect to linearly independent matrices $J_1, J_2, J_3$
we get the system of determining equations for the functions $a_k,\ b_k,\
c_k,\ d_k$
\begin{eqnarray*}
&&d_1=0,\quad d_3=0,\\
&&d_2'-Vd_1=0,\quad d_1'+Vd_2 - c_3=0,\quad d_3'-c_1=0,\\
&&c_2'-Vc_1=0,\quad c_1'+Vc_2 - b_3=0,\quad d_3'-b_1=0,\\
&&b_2'-Vb_1=0,\quad b_1'+Vb_2 - a_3=0,\quad c_3'-a_1=0,\\
&&a_2'-Va_1=0,\quad a_1'+Va_2=0,\quad a_3'=0.
\end{eqnarray*}

Integrating the above system of ODEs yields
\begin{eqnarray*}
&&d_1=0,\quad d_2=C_1,\quad d_3=0,\\
&&c_1=0,\quad c_2=C_2,\quad c_3= C_1V,\\
&&b_1=C_1V',\quad b_2=\frac{1}{2}C_1V^2 + C_3, \quad b_3=C_2V,\\
&&a_1=C_2V',\quad a_2=\frac{1}{2}C_2V^2 + C_4,\quad
a_3=C_1(V'' + \frac{1}{2}V^3) +C_3 V,
\end{eqnarray*}
where $C_1, C_2, C_3, C_4$ are arbitrary constants and furthermore
the potential $V(x)$ has to satisfy the following nonlinear ODEs
\[
C_1(V''' + \frac{3}{2} V^2V') + C_3V'=0,\quad
C_2(V'' + \frac{1}{2} V^3) + C_4V=0.
\]

Thus we have established that if the function $V(x)$ is a solution
of the stationary mKdV equation
\begin{equation}
\label{mkdv}
C_1(V''' + \frac{3}{2} V^2V') + C_3V'=0,
\end{equation}
then the initial system of ODEs (\ref{ode2}) admits the Lie symmetry
\begin{eqnarray*}
Q&=&C_1J_2\,m^3 + C_1VJ_3\,m^2 + C_1V'\, J_1\,m + \left(\frac{1}{2}C_1V^2+
C_3\right)J_2\,m\\
&&+ \left(C_1(V'' + \frac{1}{2}V^3) +C_3V\right)J_3.
\end{eqnarray*}

This symmetry solves the problem of integrability of system of ODEs
(\ref{ode2}) by quadratures due to the assertion given below.
\begin{lemma} Let the system of ODEs
\begin{equation}
\label{n4}
{\cal L}\psi\equiv \left({d\over dx} + f_a(x)Q_a\right)\psi=0,
\end{equation}
where $Q_1, Q_2, Q_3$ are constant matrices forming a basis
of the Lie algebra $so(3)$, admit a Lie symmetry
\[
X=\sum_{a=1}^3\, g_a(x)Q_a.
\]
Then it is integrable by quadratures.
\end{lemma}
{\bf Proof}.\ Making a change of dependent variables
\[
\psi\to \tilde \psi = {\cal V}(x)\psi,\quad
{\cal V}(x)=\exp\left\{\sum_{a=1}^3\, h_a(x)Q_a\right\}
\]
we can always transform the operator $X$ to become
\[
\tilde X={\cal V}^{-1}X{\cal V}=g(x)Q_1,\quad g(x)\ne 0
\]
and what is more this transformation preserves the structure of system
(\ref{n4}). The invariance criterion $[\tilde {\cal L},\ \tilde X]=0$,
where
\[
\tilde{\cal L}={\cal V}^{-1}{\cal L}{\cal V}=
\sum_{a=1}^3\, \tilde f_a(x)Q_a,
\]
implies that
\[
g(\tilde f_2Q_3 - \tilde f_3Q_2) - g'Q_1=0.
\]
As the matrices $Q_1, Q_2, Q_3$ are linearly independent, hence it
follows that
\[
\tilde f_2=0,\quad \tilde f_3 =0,\quad g={\rm const}.
\]

Consequently, the transformed system of ODEs necessarily takes the form
\[
\left({d\over dx} + \tilde f_1Q_1\right)\psi=0
\]
and is evidently integrable by quadratures. The lemma is proved.

Hence, we get a remarkable fact: {\em if $V(x)$ is a solution
of the stationary mKdV equation (\ref{mkdv}), then system of
ODEs (\ref{ode2}) is integrable by quadratures}.

Now we turn to the case when a Lie symmetry is looked for as a polynomial
in $m$ of an arbitrary order $n$
\[
Q=\sum_{k=0}^n\,\sum_{a=1}^3f_a^k(x)J_a\,m^{n-k}.
\]

The invariance criterion $[{\cal L},\ Q]=0$ yields the following
system of determining equations for the coefficients of the operator
$Q$:
\begin{eqnarray*}
&&f_1^0=0,\quad f_3^0=0,\\
&&(f_3^k)' - f_1^{k+1}=0,\quad (f_1^k)' - Vf_2^k=0,\\
&&(f_2^k)' + Vf_1^k - f_3^{k+1}=0,\quad k=0,\ldots,n-1,\\
&&(f_0^n)'=0,\quad (f_1^n)' - Vf_2^n=0,\quad (f_1^n)' + Vf_2^n=0.
\end{eqnarray*}

We have obtained two classes of solutions of the above system
of ODEs which are given below
\begin{eqnarray}
&1.& n=2N+1,\quad N\in {\bf N},\nonumber\\
&&f_1^0=0,\quad f_2^0=1, f_3^0=0,\nonumber\\
&&f_1^{2k+1}=f_2^{2k+1}=0,\quad f_3^{k+1}=R_k,\quad k=1,\ldots,N,\nonumber\\
&&f_1^{2k+2}=D_x\,R_k,\quad f_1^{2k+1}=(V-D_x^{-1}V')\,R_k,\nonumber\\
&&f_3^{2k+2}=0,\quad k=0,\ldots,N-1\nonumber
\end{eqnarray}
and the equation
\begin{equation}
\label{eq1}
D_xR_N=0
\end{equation}
holds.
\begin{eqnarray}
&2.&n=2N+2,\quad N\in {\bf N}\nonumber\\
&&f_1^0=0,\quad f_2^0=1, f_3^0=0,\nonumber\\
&&f_1^{2k+1}=f_2^{2k+1}=0,\quad f_3^{k+1}=R_k,\quad k=1,\ldots,N,\nonumber\\
&&f_1^{2k+2}=D_x\,R_k,\quad f_1^{2k+1}=(V-D_x^{-1}V')\,R_k,\nonumber\\
&&f_3^{2k+2}=0,\quad k=0,\ldots,N\nonumber
\end{eqnarray}
and the equation
\[
R_{N+1}=0
\]
holds.

In the above formulae we make use of the following notations
\[
R_k=\sum_{j=0}^k\,C_j\,(D_x^2 + V^2 - V\,D_x^{-1}V')^j\,V,\quad k=0,\ldots N+1,
\]
where $C_0,\ldots,C_{N+1}$ are arbitrary real constants and $D_x^{-1}$ is the inverse
of $D_x$.

A reader familiar with the soliton theory will immediately recognize the
operator ${\cal X}=D_x^2 + V^2 - V\,D_x^{-1}V_x$ as the recursion operator for the mKdV
equation \cite{ibr,abl}
\[
V_t + V_{xxx} + \frac{3}{2} V^2V_{x}=0.
\]
Acting repeatedly with the operator ${\cal X}$ on the trivial
conserved density $I_0=V$ we get the whole set of conserved densities of
the mKdV equation. Next, the operator
\[
{\cal Y}=D_x{\cal X}D_x^{-1}\equiv D_x^2 + V^2 + V_xD_x^{-1}V
\]
is the second recursion operator for the mKdV equation. Its repeated action
on the trivial Lie symmetry $S_0=V_x$ yields the whole hierarchy
of the higher symmetries of the mKdV equation. Hence it follows, in particular,
that the condition (\ref{eq1}) is rewritten in the form
\begin{equation}
\label{hie}
\sum_{k=0}^{N}C_kS_k=0,\quad S_k={\cal Y}^kV'.
\end{equation}

The above equation is nothing else than the higher
stationary mKdV equation. Provided $N=1$ it reduces to
the standard stationary mKdV equation (\ref{mkdv}).

Hence, due to Lemma 1 it follows the validity of the following assertion.

\begin{theo}
Let the function $V(x)$ satisfy the higher stationary
mKdV equation (\ref{hie}) with some fixed $N$ and $C_0,\ldots,C_N$.
Then, the system of ODEs (\ref{ode2}) is integrable by quadratures.
\end{theo}

It is a common knowledge that the stationary mKdV hierarchy is reduced
to the stationary KdV hierarchy with the help of the Miura
transformation (see, e.g. \cite{abl,zah}). Furthermore, the latter are
integrated in terms of $\theta$-functions \cite{nov}. Consequently,
the system of ODEs (\ref{ode1}) is also integrable by quadratures thus
giving rise to exact solutions of the initial Dirac equation
(\ref{dirac}).

Thus symmetry analysis of a very simple reduction of the Dirac
equation (there are quite a few of much more sophisticated reductions,
see \cite{fzh}) reveals such important elements of the inverse
scattering technique for the stationary mKdV equation as the recursion
operators, infinite number of conserved densities, the hierarchy of
higher symmetries of the mKdV equation which makes the integration of
system (\ref{ode1}) very rich in results. For those involved into the
inverse scattering business this is not surprising at all, since the
above used procedure is just an inversion of the famous approach to
analysis of solitonic equations via the Lax pair \cite{lax}. More
exactly, we reduce the problem of integrating linear equation to
solving the nonlinear one. The crucial point is that the this
nonlinear differential equation can be integrated which enables us to
construct the general solution of the initial linear equation.

Now it seems reasonable to carry out systematic analysis of all
reductions of the Dirac equation by inequivalent subgroups of its
symmetry group. We believe that in this way other hierarchies of
higher solitonic equations will be obtained.  This situation is in
some analogy to symmetry reductions of the self-dual Yang-Mills
equations yielding almost all known integrable solitonic equations
(see, e.g. \cite{war}).

One more important point is that the Lie transformation groups
generated by symmetries $Q$ obtained above are not subgroups of the
maximal symmetry group $C(1,3)\otimes U(1)$ admitted by initial system
(\ref{dirac}). These symmetries correspond to conditional symmetry of
the Dirac equation (\ref{dirac}).  Let us note that conditional
symmetry of the linear and nonlinear Dirac equations was studied in
\cite{fzh,zh1,zh2}.


\begin{thebibliography}{100}
\bibitem{olv} Olver P.J., {\em Applications of Lie Groups to
    Differential Equations}, Springer, New York (1986).
\bibitem{ovs} Ovsjannikov L.V., {\em Group Analysis of Differential
    Equations}, Academic Press, New York (1982).
\bibitem{ibr} Ibragimov N.Kh., {\em Transformation Groups Applied to
    Mathematical Physics}, Reidel, Dordrecht (1985).
\bibitem{abl} Ablovitz M.J., Segur H. {\em Solitons and the Inverse
    Scattering Transform}, SIAM, Philadelphia (1981).
\bibitem{zah} Zakharov V.E., Manakov S.V., Novikov S.P. and Pitaevski
  L.P. {\em Theory of Solitons: the Inverse Scattering Method},
  Consultants Bureau, New York (1980)
\bibitem{nov} Dubrovin B.A., Matveev V.B. and Novikov S.P.
  {\em Uspekhi Matem. Nauk}, {\bf 31}, 55 (1976)
\bibitem{fzh} Fushchych W.I. and Zhdanov R.Z., {\em Symmetries and
  Exact Solutions of Nonlinear Dirac Equations}, Naukova Ukraina
  Publ., Kyiv (1992)
\bibitem{lax} Lax P., {\em Comm. Pure Appl. Math.}, {\bf 21},
  467 (1968).
\bibitem{war} Ward R.S. and Wells R.O., {\em Twistor geometry
  and Field Theory}, University Press, Cambridge (1990)
\bibitem{zh1} Fushchych W.I. and Zhdanov R.Z., {\em J. Math. Phys.},
  {\bf 32}, 3488 (1991).
\bibitem{zh2} Fushchych W.I. and Zhdanov R.Z., {\em Phys. Reports},
  {\bf 172}, 123 (1989).
\end{thebibliography}
\end{document}